\documentclass[prl,twocolumn,10pt,aps,showpacs]{revtex4-1}

\usepackage{amsmath,amssymb,amstext,amsthm,amscd,mathrsfs,eucal,bm,graphics,overpic}
\usepackage{eucal}
\usepackage{relsize}
\usepackage[usenames,dvipsnames,svgnames]{xcolor} 
\usepackage{tkz-euclide}

\usepackage[T1]{fontenc}
\usepackage{floatrow}

	\def\beq{\begin{equation}}
	\def\eeq{\end{equation}}
	\def\eref#1{(\ref{eqn:#1})}
	\def\elab#1{\label{eqn:#1}}
	
	\def\fref#1{\ref{fig:#1}}
	\def\flab#1{\label{fig:#1}}

\def\e{\mathrm{e}}

\def\beq{\begin{equation}}
\def\eeq{\end{equation}}

\def\K{\mathsf{K}}
\def\Da{\mathrm{Da}}

\def\bx{\boldsymbol{x}}

\def\bq{\boldsymbol{q}}
\def\bxi{\boldsymbol{\xi}}

\newcommand{\eqn}[1]{(\ref{eqn:#1})}
\newcommand{\lab}[1]{\label{eqn:#1}}
\newcommand{\inter}[1]{\quad \textrm{#1} \quad}

\def\XXint#1#2#3{{\setbox0=\hbox{$#1{#2#3}{\int}$}
\vcenter{\hbox{$#2#3$}}\kern-.5\wd0}}

\usepackage[svgnames]{xcolor}

\definecolor{lightblue}{RGB}{11.0080,132.6080,199.6800}
\definecolor{yellow}{RGB}{191.7440,191.7440,0}
\definecolor{red}{RGB}{216.8320,41.2160,0}
\definecolor{ultramarine}{RGB}{0,32,96}

\newcommand*{\redsquare}{\textcolor{red}{\square}}
\newcommand*{\yellowdiamond}{\textcolor{yellow}{\diamond}}
\newcommand*{\lightbluecirc}{\textcolor{lightblue}{\circ}}

\def\strutdepth{\dp\strutbox}
\def\nw#1{\strut\vadjust{\kern-\strutdepth\vtop to0pt{\vss\hbox to\hsize {\hskip\hsize\hskip5pt\color{blue}$\leftarrow$\hss\strut}}}{\em \color{blue} #1}}

\usepackage[normalem]{ulem}
\begin{document}
	 
\title{\bf Dispersion in rectangular networks: \\ effective diffusivity and large-deviation rate function}

\author{Alexandra Tzella${}^{1}$ and Jacques Vanneste${}^{2}$}
\affiliation {${}^1$School of Mathematics,
University of Birmingham, Birmingham, B15 2TT, United Kingdom \\
${}^2$School of Mathematics and Maxwell Institute for Mathematical Sciences, University of Edinburgh, Edinburgh, EH9 3FD, United Kingdom}
\date{\today}
\begin{abstract}
	
The dispersion of a diffusive scalar in a fluid flowing through a network has many applications including to biological flows, porous media, water supply and urban pollution. Motivated by this, we develop a large-deviation theory that predicts the evolution of the concentration of a scalar released in a rectangular network in the limit of large time $t \gg 1$. This theory provides an approximation for the concentration that remains valid for large distances from the centre of mass, specifically for distances up to $O(t)$ and thus much beyond the $O(t^{1/2})$ range where a standard Gaussian approximation holds. A byproduct of the approach is a closed-form expression for the effective diffusivity tensor that governs this Gaussian approximation. Monte Carlo simulations of Brownian particles confirm the large-deviation results and demonstrate their effectiveness in describing the scalar distribution when $t$ is only moderately large.

\end{abstract}
\pacs{05.40.-a, 05.60.Cd, 47.51.+a, 47.56.+r, 47.85.lk}
\maketitle

In this Letter, we investigate the dispersion of a diffusive scalar released in a fluid flowing through a  rectangular network (see Fig.\ \fref{sketch}). A vivid example of application -- and a motivation for our work -- is the spreading of a pollutant released suddenly in the streets of a city with a regular grid plan such as Manhattan. The primary question concerns the form taken by the scalar concentration $C(\bx,t)$  long after release, when the disparity between the (large) scale of the scalar patch and the (small) scale of the network makes the problem challenging. The question arises in numerous applications across science and engineering besides urban pollution \cite{Belcher2005,Belcher_etal2015}: vascular and respiratory flows \cite{Truskey_etal2004}, microfluidic devices \cite{Thorsen_etal2002,Stone_etal2004}, porous media \cite{Adler1992,BrennerEdwards1993,Sahimi1993} and water distribution \cite{LiouKroon1987} for example. 
Its answer sheds light on the subtle interplay between advection, diffusion and geometry that controls dispersion in networks.

As is typical for advection--diffusion problems, $C(\bx,t)$ for $t \gg 1$ can be approximated  by a Gaussian, parameterised by an effective diffusivity tensor \cite{Taylor1953,MajdaKramer1999}. This approximation applies only to the core of the scalar distribution, specifically to distances $O(t^{1/2})$ away from the centre of mass: the network geometry leads to non-Gaussian behaviour in the tails of the distribution. 
These tails are important in applications  where low concentrations are  critical, e.g., for highly toxic chemicals or in the presence of amplifying chemical reactions. 
To capture  both the Gaussian core and the tails, we develop a large-deviation theory \cite{FreidlinWentzell1984,Freidlin1985} 
that leads to  a general approximation, of the form   $C(\bx,t) \propto \exp(-t g(\bx/t))$ and holds for distances up to $O(t)$ away from the centre of mass \cite{HaynesVanneste2014a}. Here $g$ is a rate function which we compute  (by solving a transcendental equation) and approximate explicitly in asymptotic limits. Its quadratic approximation gives a closed-form expression for the effective diffusivity controlling the core of the scalar distribution. Monte Carlo simulations confirm the effective-diffusivity and large-deviation results and demonstrate the benefits of the latter, particularly for moderately large $t$ when the non-Gaussian behaviour is most conspicuous. 

\begin{figure}[b]
	\begin{center}
	  \includegraphics[width=.85\linewidth]{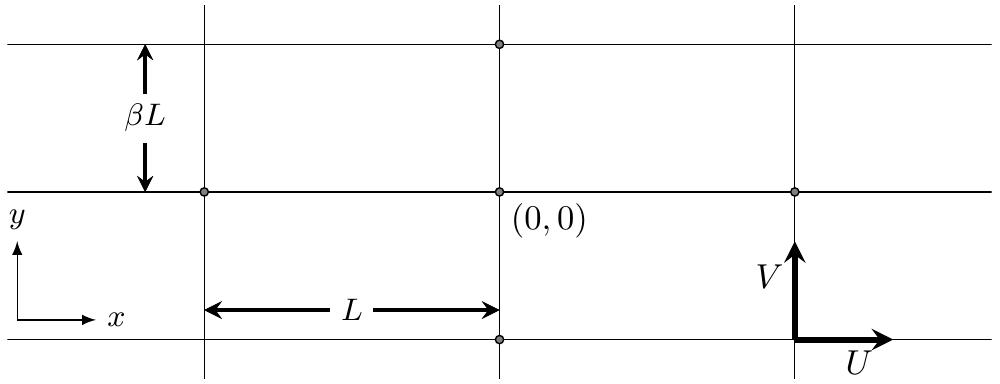}
\end{center}
\caption{
A section of the
rectangular network which includes the vertex at $(x,y)=(0,0)$. Fluid flows with velocity %$(U,V)$
 $U$ and $V$ along edges of  length  $L$ and $\beta L$.  
 }
\flab{sketch}
\end{figure}

{\it Model.}---
We consider the rectangular network in Fig.\ \fref{sketch} composed of one-dimensional edges 
of length $L$ and  $\beta L$ in the $x$- and $y$-directions 
along  
which fluid flows with uniform velocity
$U$ and $V$.  
This simple model has proved remarkably effective in describing pollution spreading through dense city centres \cite{Belcher2005,Belcher_etal2015}. More broadly, it  provides an excellent prototype for geometry-induced non-Gaussianity and its description by large deviations.

Taking $L$ as reference length and  $L^2/\kappa$ as reference time, the one-dimensional advection--diffusion equations for the scalar concentration $C$  read
\beq \elab{ad-dif}
\partial_t C + U \partial_x C = \partial_{xx}^2 C \inter{and}
\partial_t C + V \partial_y C = \partial_{yy}^2 C,
\eeq
in edges oriented  along $x$ and $y$  \footnote{For pipe  or channel flows, $(U,V)$ can be interpreted as a section-averaged velocity and $\kappa$ as a Taylor diffusivity.}. The non-dimensional parameters $U$ and $V$ are  P\'eclet numbers measuring the strength of advection relative to diffusion.
These equations are supplemented by boundary conditions applied at the vertices separated by distances $1$ in  $x$ and $\beta$  in  $y$.
The boundary conditions express (i) continuity of $C$,
\beq \elab{cont}
C\rvert_\mathrm{W} = C|_\mathrm{E} = C|_\mathrm{S} = C|_\mathrm{N},
\eeq
where the subscripts denote the limiting value to the west, east, etc.\ of the vertex, and (ii)  vanishing of the net concentration flux which, on using \eref{cont}, simplifies into
\beq \elab{flux}
\partial_x C |_\mathrm{W} +  \partial_y C |_\mathrm{S}= \partial_x C |_\mathrm{E} + \partial_y C |_\mathrm{N}.
\eeq
Eqs.\ \eqn{ad-dif}--\eqn{flux} form a closed system which can  
be solved numerically to predict the evolution of $C$ for arbitrary initial conditions 
(e.g. using Laplace transforms \cite{deArcangelis_etal1986, Koplik_etal1988, Heaton_etal2012}).
Here we consider a scalar  initially released at a vertex  taken
to be   the origin so that $C(x,y,0)=\delta(x) \delta(y)$.

{\it Large deviations.}---%
Analytic progress is possible using the theory of large deviations \cite{FreidlinWentzell1984,Freidlin1985}. This  describes the  concentration in the long-time limit $t \gg 1$ as \cite{HaynesVanneste2014a} 
\beq \elab{largedevi}
C \sim t^{-1} \phi(x,y) \e^{-t g(\bxi)}, \quad \textrm{with} \ \ \bxi =(x,y)/t \; \in \mathbb{R}^2.
\eeq
The rate (or Cram\'er) function $g(\bxi)$ provides a continuous approximation for the most rapid changes in $C$ and is the main object of interest.  
 The function $\phi$ is supported on the network and has periods $1$ and $\beta$ in $x$  and $y$. The factor $t^{-1}$ is imposed by normalisation. Introducing \eqn{largedevi} into \eqn{ad-dif} leads to
 \setcounter{footnote}{100}\footnote{See appended  Supplemental Material  for details of the derivation.}
\begin{eqnarray}
 \partial_{xx} \phi - (U + 2 q_x) \partial_x \phi + (U q_x+q_x^2) \phi &=& f(\bq) \phi, \lab{eig1} \\
\partial_{yy} \phi - (V + 2 q_y) \partial_y \phi + (V q_y+q_y^2) \phi &=& f(\bq) \phi. \lab{eig2}
\end{eqnarray}
To write these we have defined
\beq\elab{Legendre}
\bq = (q_x,q_y)^\mathrm{T} = \nabla_{\bxi} g \inter{and} f(\bq) = \bxi \cdot \bq - g(\bxi),
\eeq
which implies that $f$ and $g$ are Legendre transforms of one another, with $\bq$ and $\bxi$ the dual independent variables. Eqs.\ \eqn{eig1}--\eqn{eig2} are supplemented by the boundary conditions
inferred from \eqn{cont} and \eqn{flux}:  continuity 
of $\phi$  
and
\beq \lab{flux2}
\partial_x \phi|_\mathrm{W} + \partial_y \phi|_\mathrm{S} =
\partial_x \phi|_\mathrm{E} +  \partial_y \phi|_\mathrm{N}.
\eeq
Together, \eqn{eig1}--\eqn{flux2}
form a family of eigenvalue problems parameterized by $\bq$, with $f(\bq)$ as the eigenvalue.

We solve Eqs. \eqn{eig1}--\eqn{eig2}  to obtain explicit expressions for the eigenfunction $\phi$ in the
$4$ edges incident to the vertex $(0,0)$ using periodicity \cite{Note101}.
Introducing the solution into  \eref{flux2}  gives
\beq \lab{trans}
\begin{split}
&\frac{\alpha_U  \cosh \alpha_U}{\sinh \alpha_U} + \frac{\alpha_V \cosh(\alpha_V \beta) }{\sinh(\alpha_V \beta)} = \\
&\frac{\alpha_U \cosh (q_x + U/2)}{\sinh \alpha_U}  + \frac{\alpha_V  \cosh((q_y + V/2) \beta)}{\sinh(\alpha_V \beta)},
\end{split}
\eeq
where $\alpha_U = \sqrt{f(\bq)+U^2/4}$ and similarly for $\alpha_V$.
This transcendental equation for $f(\bq)$ 
is our central result. It can be solved numerically for a range of $\bq$ to obtain $f(\bq)$; the rate function $g(\bxi)$ is deduced by  Legendre transform.   We start our analysis by considering the behaviour of $g(\bxi)$ near its minimum. This provides a closed-form expression for the effective diffusivity of the network.

\begin{figure} 
\begin{center}
\includegraphics[width=.95\linewidth]{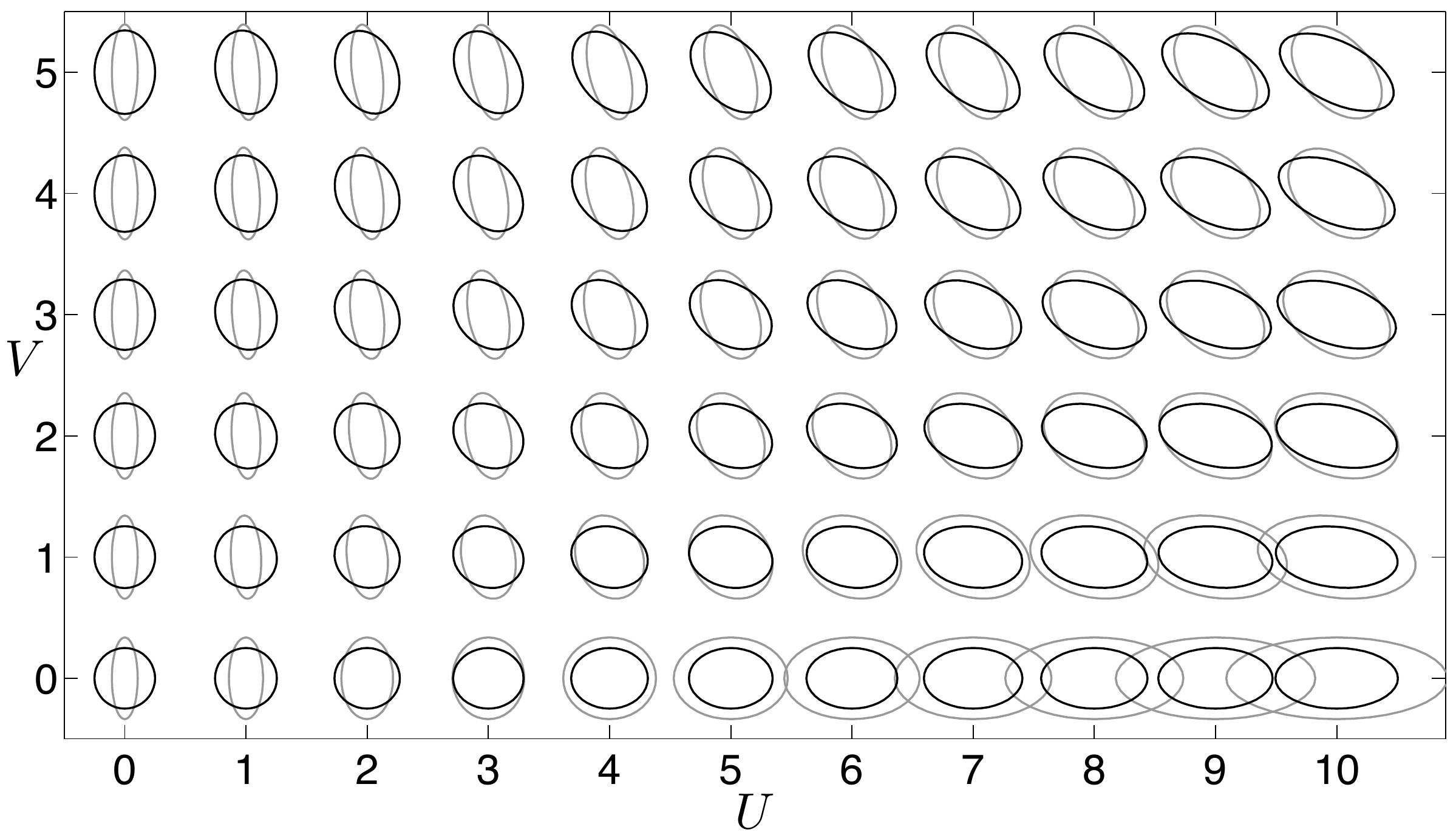}
\caption{Ellipses of constant $\bx^\mathrm{T} \mathsf{K}^{-1} \bx$ representing the effective diffusivity tensor $\mathsf{K}$ as a function of $U$ and $V$ for $\beta=1$ (black) and $\beta=10$ (grey).} \flab{ellipse}
\end{center}
\end{figure}

{\it Effective diffusivity.}---%
The Gaussian, diffusive approximation
\beq\elab{Gaussian}
C(\bx,t) \sim t^{-1} \e^{-(\bx - \bxi_* t)^\mathrm{T} \mathsf{K}^{-1}  (\bx - \bxi_* t)/(4t)},
\eeq
is deduced from \eqn{largedevi} by Taylor expanding $g(\bxi)$ around its minimum $\bxi_*$, identified as the velocity of the centre of mass of the scalar.
It can be shown \cite{Note101} that $\bxi_* = \nabla_{\bq} f(\bm{0})$, and that the effective diffusivity tensor is $\K = \nabla_{\bq} \nabla_{\bq} f(\bm{0})/2$, i.e.,\ half the Hessian of $f$ at $\bq=\bm{0}$.
Introducing the Taylor expansion of $f$ around $\bm{q}=\bm{0}$ into (9) and solving gives, after lengthy manipulations,
\beq \elab{meanvel}
\bxi_* = \left(\frac{U}{1+\beta} , \frac{\beta V}{1+\beta} \right)
\eeq
and the components  
\begin{eqnarray}
\K_{11} &=& \frac{(1+\beta)^2+\beta^2 U^2 \left(h(U)+\beta h(\beta V)\right)}{(1+\beta)^3}, \elab{k11} \\
\K_{22} &=& \frac{\beta (1+\beta)^2+\beta^2 V^2 \left(h(U)+\beta h(\beta V)\right)}{(1+\beta)^3}, \elab{k22} \\
\K_{12} &=& \K_{21} = -  \frac{\beta^2 U V \left(h(U)+\beta h(\beta V)\right)}{(1+\beta)^3}, \elab{k12}
\end{eqnarray}
of $\K$, where $h(x)=x^{-2} (x \coth(x/2)/2-1)$. Note that effective diffusivities are more commonly derived using 
 homogenization \cite{Bensoussan_etal1978,RubisteinMauri1986,Mei1992,AuriaultAdler1995} or the method of moments \cite{Aris1956,Brenner1980}: solving their cell problem amounts to a perturbative solution of \eqn{eig1}--\eqn{flux2} \cite{HaynesVanneste2014a}.

The explicit expressions \eqn{meanvel}--\eqn{k12}
illustrate the complex interplay between advection and diffusion that determines 
 dispersion in networks.
They are visualised for a range of $U$ and $V$ and two values of $\beta$ as ellipses of constant $\bx^\mathrm{T} {\mathsf{K}}^{-1} \bx$ (corresponding to constant concentration) in Fig.\ \fref{ellipse}.
For   $U,\,V\ll 1$  (small P\'eclet number), the asymptotic formula
$h(x) = 1/12 +O( x^2)$ as $x\to 0$ provides the approximation  $\K_{11} \sim 1/(1+\beta) + \gamma U^2$,
$\K_{22} \sim\beta/(1+\beta)+\gamma V^2$ and $\K_{12} \sim - \gamma UV$, with $\gamma=\beta^2/(12(1+\beta)^2)$.

For $U,\,V \gg 1$   (large P\'eclet number), we use $h(x) = 1/(2 x)+O(x^{-2})$ as $x \to \infty$ to approximate the effective diffusivity components as
$\K_{11} \sim \delta U^2$, $\K_{22}\sim \delta V^2 $ and $\K_{12} \sim -\delta UV$, with $\delta=\beta^2(U^{-1} +  V^{-1})/(2(1+\beta)^{3})$.
These grow linearly in $U$ and $V$ which dimensionally corresponds to components 
that are independent of the molecular diffusivity $\kappa$. 
This is characteristic of a regime termed 
geometric \cite{BouchaudGeorges1990} or mechanical \cite{KochBrady1985,Sahimi1993} dispersion.
The tensor $\mathsf{K}$
is singular to leading order in $U$ and $V$: effective diffusion is strong in the direction $(-U,V)$ but weak in the perpendicular direction $(V,U)$ 
(see  Fig.\ \fref{ellipse}) 
For $U \gg 1$ and $V \ll 1$, i.e., strong flow in the $x$-direction, $\K_{11} \sim \beta^3 U^2 /(12(1+\beta)^3) \gg \K_{22},\, \K_{12}$. This corresponds to a mostly longitudinal diffusivity with a $\kappa^{-1}$ scaling characteristic of   Taylor dispersion  \cite{Taylor1953}. 
%\nw{The scaling behaviour of both regimes   can be   heuristically obtained  by considering a  spatially discrete random-walk model in which particles move between  neighbouring nodes at a characteristic  time whose value is 
%\cite{Koplik_etal1988}.}
Note that the geometric and Taylor regimes can be understood in terms of a random-walk model with correlation time determined by advection in the first case and molecular diffusion in the second   \cite{Koplik_etal1988}.

{\it Rate function.}---Effective diffusivity provides a partial description of dispersion: the  rate function $g$ obtained from \eref{trans} is much more informative. This is demonstrated in Fig.\
\fref{g_geff} which shows typical examples of $g$ obtained numerically for two values of $(U,V)$ (for $\beta=1$)
and its quadratic approximation corresponding to the Gaussian \eref{Gaussian}.
This   approximation is excellent in the vicinity of $\bxi_*$, with circular (Fig. \fref{g_geff}(a)) and elliptical contours (Fig.\ \fref{g_geff}(b)).
Beyond the vicinity of $\bxi_*$, it is inadequate,  failing for instance to capture the anisotropy of $g$ and hence of $C$ for $U=V=0$, or underestimating $g$ in large portions of the $\bxi$-plane (hence overestimating $C$ by an exponentially large factor) for $U=V\not=0$.   
Note however that for $U=V$ and $\beta=1$, $f$ is exactly quadratic along the line $q_x=q_y$ (see \eref{trans});   hence, $g$  coincides with its quadratic approximation for $\xi_x=\xi_y$, as  evident in Fig.\ \fref{g_geff}.

\begin{figure}[t]
\begin{center}
\includegraphics[width=.99\linewidth]{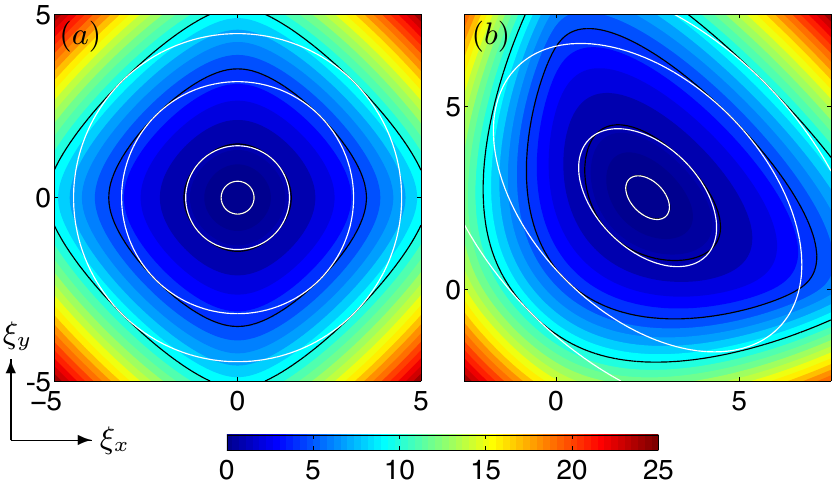}
\end{center}
% 	\begin{center}
% %
% \begin{overpic}[width=.95\linewidth]%
%       {g_geff_panel.pdf}
% \put(-2,5){\vector(1,0){10}}
% \put(-2,5){\vector(0,1){10}}
% \put(-3,17){$\xi_y$}
% \put(9,4){$\xi_x$}
% \put(4,54){$(a)$}
% \put(55,54){$(b)$}
% \end{overpic}
% \end{center}
%
\caption{
(Color online.)
Rate function $g$ calculated numerically from \eqn{trans}
for $\beta=1$ and (a) $(U,V)=(0,0)$ and (b)$\,(5,5)$.
Selected contours  (with values 0.1, 1, 5 and 10) compare  $g$ (black) with its quadratic, Gaussian approximation (white)
 \eref{Gaussian}.
This approximation is clearly valid near the minimum  $\bxi_\ast$ of $g$.
 }
\flab{g_geff}
\end{figure}

\begin{figure}[h]
	\begin{center}
	  \includegraphics[width=.99\linewidth]{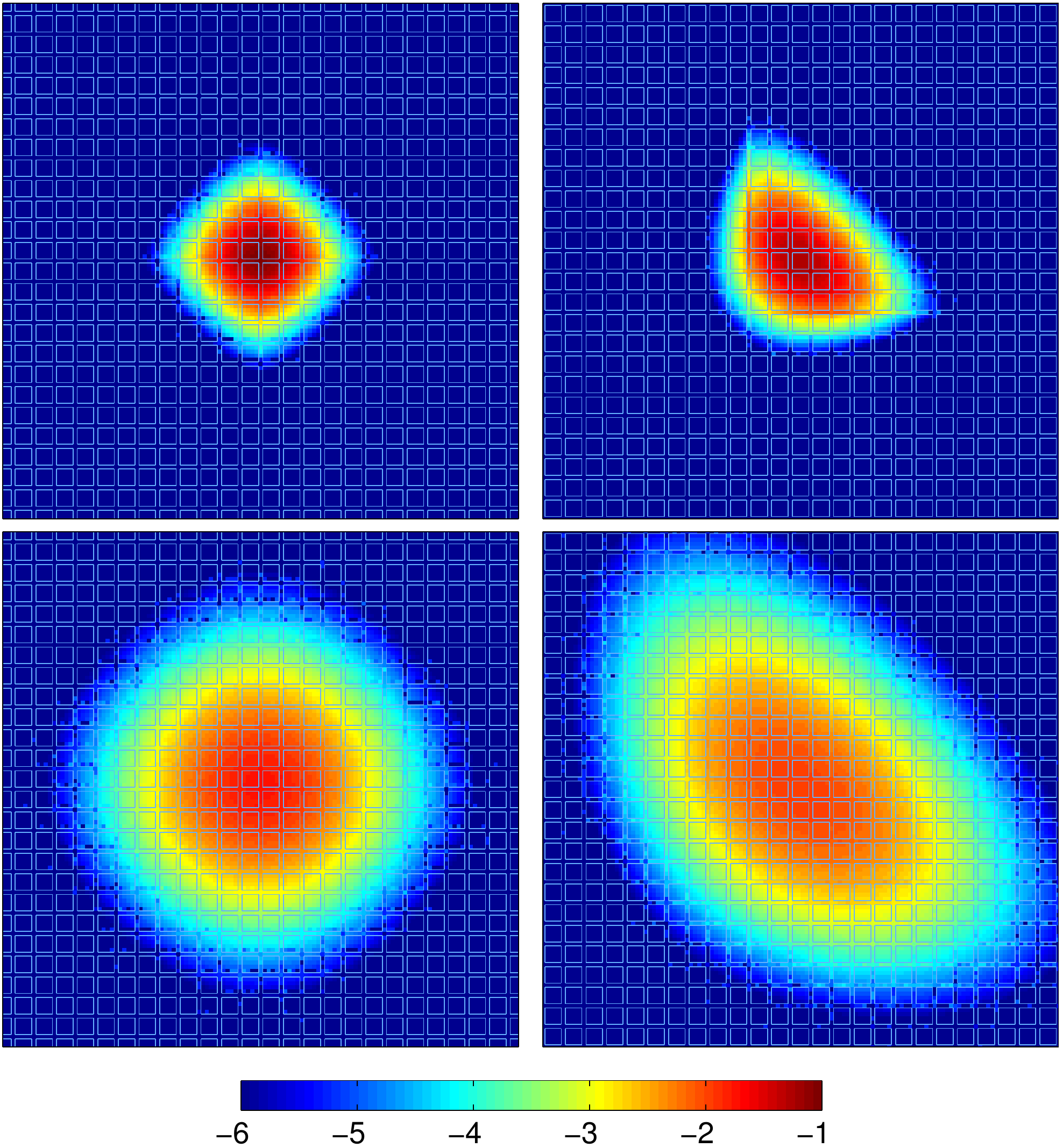}
\end{center}
% \begin{center}
%  \begin{overpic}[width=1\linewidth]{pdf-noresample_g_panelb_t1_5_lightblue.pdf}
%  \put(2,95){\textcolor{white}{$(a)$}}
%  \put(48,95){\textcolor{white}{$(b)$}}
%  \end{overpic}
%  \end{center}
 \vspace{-.5cm}
 \caption{(Color online.) Snapshots of $\log_{10}C$
  for $t=1$ (top) and $t=5$ (bottom) for the  parameters of Fig. \fref{g_geff}.
Numerical results are shown inside the network.
The large-deviation prediction \eref{largedevi} and  \eref{trans} is shown outside the network.}
 \flab{snapshots}
\end{figure}

The limitations of the quadratic (Gaussian) approximation are best demonstrated by considering the large-$\bxi$ behaviour of $g(\bxi)$ or, equivalently, the large-$\bq$ behaviour of $f(\bq)$.
In this regime, and with the distinguished scaling $U,\, V = O(|\bq|)$, \eqn{trans} reduces to
\beq \elab{largeq}
\begin{split}
\alpha_U + \alpha_V \sim \alpha_U &\cosh(q_x+U/2)\e^{- \alpha_U}\\
&+ \alpha_V \cosh(\beta(q_y+V/2))\e^{- \beta \alpha_V}.
 \end{split}
\eeq
Either term on the right-hand side is exponentially large, precluding the solution of \eref{largeq}  unless
\beq\elab{f-largeq}
f(\bq) \sim \max(q_x^2 + U q_x,q_y^2 + V q_y).
\eeq
This gives a leading-order approximation to $f$ which, remarkably, is independent of $\beta$. The Legendre transform of \eref{f-largeq} is cumbersome for arbitrary $U$ and $V$, %\nw{\cite{Note101}}, %\cite{supp},
but physical insight is gained by considering limiting cases.
For  $U=V=0$, \eref{f-largeq} leads to
$g(\bxi) \sim \left( | \xi_x| + |\xi_y| \right)^2/4$,
in accordance with the diamond-shaped contours of $g$ for large $\bxi$ in Fig.\ \fref{g_geff}(a).
This implies a concentration $C \sim \exp (-(|x|+|y|)^2/(4t))$, which can be interpreted as  a generalised form of diffusion   with the Euclidian distance  replaced by the $L^1$ (or Manhattan) distance.
 When $U q_x$ and $V q_y$ dominate in \eref{f-largeq}, the linear dependence of $f$ on $\bq$ implies that $g \to \infty$ as $\xi_x \to U$ and as $\xi_y \to V$, reflecting the finite propagation speed of the scalar when molecular diffusion is neglected against advection.

The large-P\'eclet regime $U,\, V \gg 1$ (and $\bxi=O(1)$) is of interest. In this regime, $f=O(U,V)$ and \eqn{trans} becomes
\beq \elab{largePe}
U \left( \e^{q_x - f/U} -1\right) + V\left(\e^{\beta(q_y - f/V)}-1\right)=0,
\eeq
which implies a concentration  
independent of molecular diffusivity, generalising the notion of geometric or mechanical dispersion to the large-deviation regime.

{\it Monte Carlo simulations.}---%
We now test our predictions against
Monte Carlo simulations of Brownian particles.
The concentration, derived as the probability density function (PDF) of their positions $\bm{X}(t)$, is compared with the large-deviation estimate in Fig. \fref{snapshots}.
The PDF is obtained from an ensemble of $N=10^6$   particles by integrating the
 stochastic differential equations  associated with \eref{ad-dif}, with the additional microscopic rule that  particles entering a vertex exit through a random edge 
\footnote{
The details of the microscopic rule, in particular, the probability of exit by each of the four edges, do not affect the form of the macroscopic equations.}.
Although formally valid for $t\gg 1$, 
the large-deviation approximation \eref{largedevi}  
 is remarkably accurate for the moderate values of $t=1$ and $5$ considered. 
 Its relevance is clear at $t=1$, when comparing Figs.\ \fref{g_geff}  and \fref{snapshots}.
As time progresses, the Gaussian approximation \eref{Gaussian} becomes sufficient to describe the bulk of the scalar patch which assumes a characteristic elliptical form 
(cf. Fig. \fref{g_geff}).

\begin{figure}[t]
		\begin{center}
		  \includegraphics[width=.99\linewidth]{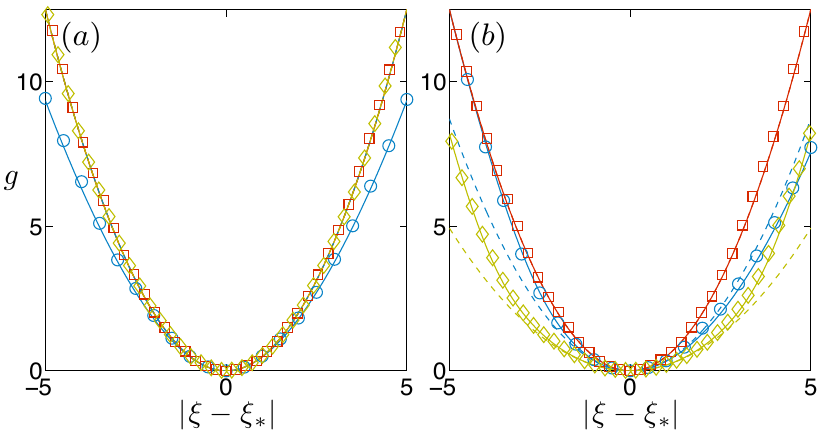}
	\end{center}
	% \begin{overpic}[width=.96\linewidth]{g_importancesampling_panel_colour.pdf}
% 		 \put(7,48){$(a)$}
% 		 \put(57,48){$(b)$}
%      \end{overpic}
 \caption{(Color online.)
 Cross sections of the rate function  $g$  for the parameters of Fig. \fref{g_geff}.
 The  large-deviation and Gaussian predictions  (solid and dashed lines) are compared with
 Monte Carlo results (symbols) as a function of $|\bxi-\bxi_*|$
 in the directions $(1,0)$ ($\lightbluecirc$), $(1,1)$ ({\tiny$\redsquare$})  
 and $(1,-1)$ ($\yellowdiamond$).  Because $U=V$, the Gaussian and large-deviation predictions coincide in the  direction $(1,1)$ (and in direction $(1,-1)$ in panel (a)).}
 \flab{g-MC}
\end{figure}

A  detailed assessment of the large-deviation approximation requires a careful numerical evaluation of the rate function $g$. This is achieved by estimating its Legendre transform as the scaled cumulant generating function \cite{Ellis1995,DemboZeitouni1998,Touchette2009}
$
f(\bm{q})=\lim_{t\to\infty} t^{-1} \log \mathbb{E} \, \e^{\bm{q}\cdot\bm{X}(t)},
$
where $\mathbb{E}$ is
the expectation   over the Brownian motion.
To reduce sampling error to an acceptable level,
we have adopted the pruning--cloning technique described in  \cite{HaynesVanneste2014a} based on \cite{Grassberger1997}. 
Fig.\ \fref{g-MC} shows an excellent agreement between large-deviation predictions and numerical results (obtained   for $N=10^3$ particles at $t=5$) and illustrates the restricted range of validity of the Gaussian approximation.

{\it Conclusion.}---
We characterise the dispersive properties of a rectangular network by a 
rate function $g$ deduced from \eref{trans}. This describes the scalar concentration over a broad range of distances $|\bx-\bxi_* t|=O(t)$ which proves particularly pertinent for moderately long times. 
In the narrower range $|\bx-\bxi_* t|=O(t^{1/2})$, it recovers the Gaussian, diffusive approximation  and provides a convenient route to derive
 the  effective diffusivity. Several conclusions can be drawn from the results: (i) in the absence of advection, the dispersion   switches from a standard, $L^2$ diffusion with diffusivity $\kappa/2$ near the point of release, to an $L^1$ diffusion with diffusivity $\kappa$ at large distances; (ii) correspondingly, the Gaussian approximation misrepresents the shape of the scalar patch   and underestimates its area (by a factor $\pi/4$); 
(iii) advection leads to a complex, anisotropic behaviour, even in the Gaussian regime, with an  enhancement of dispersion in the direction $(-U,V)$ corresponding to a constant advective travel time $x/U + y/V$;   (iv) strong advection (large P\'eclet number) leads to a geometric-dispersion regime, in which the  rate function, and hence the effective diffusivity, are independent of the molecular diffusivity $\kappa$; 
(v) advection aligned with one of the axes of the network is anomalous in this respect, with an effective diffusivity that instead scales like $\kappa^{-1}$ as in Taylor dispersion; (vi) the Gaussian approximation can under- and overpredict the scalar concentration for $|\bx|=O(t)$, depending on $\bx$, by a factor that is exponentially large in $t$.

We emphasise that our large-deviation approach generalises straightforwardly to other periodic networks. It can capture anomalous diffusion \cite{BouchaudGeorges1990} (when $g$ is not quadratic near its minimum) and  be further extended to fractal and random networks \cite{deArcangelis_etal1986,Koplik_etal1988, ben-AvrahamHavlin2000,Kang_etal2011}. 
Our results are also directly applicable to reactive fronts: a Fisher-Kolmogorov-Petrovskii-Piskunov (FKPP) reaction, which adds $\Da \, C (1-C)$ to \eqn{ad-dif} (with the Damk\"ohler number  $\Da$ as non-dimensional reaction rate), leads to the emergence of a concentration front. Its long-time speed of propagation $v$ is determined by $g$ through the condition $g(v) = \Da$ \cite{Freidlin1985,TzellaVanneste2015}.

\smallskip
The work was  supported by EPSRC (Grant No.\ EP/I028072/1).

%\bibliography{networks}

%merlin.mbs apsrev4-1.bst 2010-07-25 4.21a (PWD, AO, DPC) hacked
%Control: key (0)
%Control: author (8) initials jnrlst
%Control: editor formatted (1) identically to author
%Control: production of article title (-1) disabled
%Control: page (0) single
%Control: year (1) truncated
%Control: production of eprint (0) enabled
%

\clearpage

\widetext

% \begin{center}
% \textbf{\textsc{Supplemental Material}}
% \end{center}

\section{Supplemental Note}
	
In this supplemental note, we provide details of the derivation of the eigenvalue problem determining the rate function $g$. We also deduce the corresponding effective diffusivity.

\subsection{Eigenvalue problem}
As discussed in the Letter, in the large-deviation regime, the concentration $C$ of a dispersing scalar takes the asymptotic form
\beq \lab{largedevi}
C \sim t^{-1} \phi(x,y) \e^{-t g(\bxi)}, \quad \textrm{with} \ \ \bxi =(\xi_x,\xi_y)=(x,y)/t\; \in \mathbb{R}^2.
\eeq
Substituting \eqn{largedevi} into the advection--diffusion equations 
%$\partial_t C + U \partial_x C = \partial_{xx}^2 C$ and $\partial_t C + V \partial_y C = \partial_{yy}^2 $
\beq \elab{ad-dif}
\partial_t C + U \partial_x C = \partial_{xx}^2 C \inter{and}
\partial_t C + V \partial_y C = \partial_{yy}^2 C,
\eeq
and equating powers of $t^{-1}$ yields, at leading order, 
\begin{eqnarray}
\partial_{xx} \phi - (U + 2 \partial_{\xi_x} g) \partial_x \phi + (U \partial_{\xi_x} g+(\partial_{\xi_x} g)^2) \phi &=& (\bxi \cdot \nabla_{\bxi} g - g) \phi,\lab{eig1a} \\
\partial_{yy} \phi - (V + 2 \partial_{\xi_y} g) \partial_y \phi + (V \partial_{\xi_y} g+(\partial_{\xi_y} g)^2) \phi &=& (\bxi \cdot \nabla_{\bxi}  g - g)\phi. \lab{eig2a} 
\end{eqnarray}
Eqs.\ \eqn{eig1a}--\eqn{eig2a} are supplemented by a set of boundary conditions applied at the network's vertices that are  inferred from continuity and zero net flux of $C$ (Eqs. (2)--(3) in the Letter). 
These readily imply continuity of $\phi$ and
\[
\partial_x \phi|_\mathrm{W}-\phi|_\mathrm{W}\partial_{\xi_x}g|_\mathrm{W}  
+ \partial_y \phi|_\mathrm{S}-\phi|_\mathrm{S}\partial_{\xi_y}g|_\mathrm{S}   =  
\partial_x \phi|_\mathrm{E}-\phi|_\mathrm{E}\partial_{\xi_x}g|_\mathrm{E} 
+  \partial_y \phi|_\mathrm{N}-\phi|_\mathrm{N}\partial_{\xi_y}g|_\mathrm{N},
\]
which further simplifies into % a no flux equation for $\phi$
\beq \lab{flux2}
\partial_x \phi|_\mathrm{W}
+ \partial_y \phi|_\mathrm{S}   =
\partial_x \phi|_\mathrm{E}
+  \partial_y \phi|_\mathrm{N},
\eeq
 (Eq. (8) in the Letter) once we use continuity of $\phi$ and $\nabla_{\bm{\xi}} g$.

We let $\bq=\nabla_{\bxi} g(\bxi)$ and  $f(\bq) = \bxi \cdot \bq - g(\bxi)$ (Eq. (7) in the Letter). Treating  $\bq$ as a parameter, we obtain the eigenvalue problem 
\begin{eqnarray}
 \partial_{xx} \phi - (U + 2 q_x) \partial_x \phi + (U q_x+q_x^2) \phi &=& f(\bq) \phi, \lab{eig1} \\
\partial_{yy} \phi - (V + 2 q_y) \partial_y \phi + (V q_y+q_y^2) \phi &=& f(\bq) \phi. \lab{eig2}
\end{eqnarray}
with $f(\bq)$ as the eigenvalue (Eqs. (5)--(6) in the Letter).
The focus is on the principal eigenvalue (that with maximum real part) because it corresponds to the slowest decaying solution of \eqn{largedevi}.
The Krein--Rutman theorem implies that this eigenvalue is unique, real and isolated, with a positive associated eigenfunction $\phi>0$. Moreover, $f(\bq)\geq 0$ and is convex  so that $f(\bq)$ and $g(\bxi)$ are related by a Legendre transform 
\[
g(\bxi)=\sup_{\bq}(\bq\cdot\bxi-f(\bq))\quad\text{and}\quad f(\bq)=\sup_{\bxi}(\bq\cdot\bxi-g(\bxi))
\]
from where $\bxi=\nabla_{\bq} f(\bq)$.

We now solve \eqn{eig1}--\eqn{eig2}  explicitly. Consider the intersection at $(x,y)=(0,0)$ and denote by $\phi_\mathrm{E}(x)$ the eigenfunction in the street to the east of it, and by $A$ the value of $\phi$ at the intersection $(0,0)$ and hence, by periodicity, at all intersections. Solving \eqn{eig1} with the boundary conditions $\phi_\mathrm{E}(0)=\phi_\mathrm{E}(1)=A$, we find that
\[
\phi_\mathrm{E}(x) = \frac{A}{\sinh \alpha_U} \left( \e^{(q_x+U/2)x} \sinh(\alpha_U(1-x)) + \e^{(q_x+U/2)(x-1)} \sinh(\alpha_U x) \right),
\]
where $\alpha_U = \sqrt{f(\bq)+U^2/4}$.
The solution to the west of $(0,0)$ is found by substituting $x \mapsto x+1$ in this expression to obtain
\[
\phi_\mathrm{W}(x) = \frac{A}{\sinh \alpha_U} \left(- \e^{(q_x+U/2)(x+1)} \sinh(\alpha_U x) + \e^{(q_x+U/2)x} \sinh(\alpha_U (x+1)) \right).
\]
Similarly, the solution $\phi_\mathrm{N}$ to the north is found solving \eqn{eig2} with $\phi_\mathrm{N}(0)=\phi_\mathrm{N}(\beta)=A$ to find
\[
\phi_\mathrm{N}(y) = \frac{A}{\sinh (\alpha_V\beta)} \left( \e^{(q_y+U/2)y} \sinh(\alpha_V(\beta-y)) + \e^{(q_y+V/2)(y-\beta)} \sinh(\alpha_V y) \right),
\]
where $\alpha_V = \sqrt{f(\bq)+V^2/4}$. The substitution $y \mapsto y + \beta$ then gives
\[
\phi_\mathrm{S}(y) = \frac{A}{\sinh (\alpha_V \beta)} \left(- \e^{(q_y+V/2)(y+\beta)} \sinh(\alpha_V y) + \e^{(q_y+V/2)y} \sinh(\alpha_V (y+\beta)) \right).
\]
We can now apply the boundary condition \eqn{flux2} by evaluating the derivatives of the solution. After some simplifications, this leads to  
\beq \lab{trans}
\frac{\alpha_U  \cosh \alpha_U}{\sinh \alpha_U} + \frac{\alpha_V \cosh(\alpha_V \beta) }{\sinh(\alpha_V \beta)} = \frac{\alpha_U \cosh (q_x + U/2)}{\sinh \alpha_U}  + \frac{\alpha_V  \cosh((q_y + V/2) \beta)}{\sinh(\alpha_V \beta)}
\eeq
which is Eq. (9) in the Letter.

\subsection{Effective diffusivity}

The rate function $g$ has a single minimum, $\bxi_*$ say, around which it can be expanded according to
\beq \lab{gexp}
g(\bxi) \sim g(\bxi_*) + \tfrac{1}{2} (\bxi - \bxi_*)^\mathrm{T} \nabla_{\bxi} \nabla_{\bxi}  g(\bxi_*) (\bxi - \bxi_*),
\eeq
where $\nabla_{\bxi}  \nabla_{\bxi}  g(\bxi_*) $ is the Hessian of $g$ (matrix of second derivatives) evaluated at $\bxi_*$. It follows from the Legendre transform that $\bq = \nabla_{\bxi} g =0$ corresponds to the minimum of $g$, hence $\bxi_*=\nabla_{\bq} f(\mathbf{0})$. 
Taking the gradient of the relation $\bq = \nabla_{\bxi} g$ with respect to $\bxi$ and evaluating at $\bxi_*$ gives
\beq
\nabla_{\bxi}  \nabla_{\bxi}  g(\bxi_*) = \nabla_{\bxi} \bq (\bxi_*).
\eeq
On the other hand, the gradient with respect to $\bxi$ of $\bxi = \nabla_{\bq} f$ and the chain rule give
\beq
I = \nabla_{\bq} \nabla_{\bq} f \cdot \nabla_{\bxi} \bq,
\eeq
where $I$ is the identify matrix. Evaluating at $\bxi=\bxi_*$ and, correspondingly, $\bq=0$ leads to the standard relation between the Hessians of $g$ and $f$,
\beq
\nabla_{\bxi}  \nabla_{\bxi}  g(\bxi_*) = (\nabla_{\bq} \nabla_{\bq} f (\mathbf{0}))^{-1}.
\eeq
Introducing this into \eqn{gexp} and using in \eqn{largedevi} yields the Gaussian approximation 
\beq\lab{Gaussian}
C(\bx,t) \sim t^{-1} \e^{-(\bx - \bxi_* t)^\mathrm{T} \mathsf{K}^{-1}  (\bx - \bxi_* t)/(4t)},
\eeq
for the concentration, with the effective diffusivity tensor $\K=\nabla_{\bq} \nabla_{\bq} f (\mathbf{0})/2$. This is Eq.\ (10) of the Letter. 

In order to deduce explicit expressions for $\bxi_* = \nabla_{\bq} f(\mathbf{0})$ and $\K$ from \eqn{trans}, we introduce the expansion
\beq
f = \bxi_* \cdot \bq + \bq^\mathrm{T} \K \bq + O(|\bq|^3)
\eeq 
into \eqn{trans}, expand in powers series of $\bq$ and solve at order $O(|\bq|)$ for $\bxi_*$ and $O(|\bq|^2)$ for $\K$. This is best carried out using a symbolic-algebra package. 

% \appendix*
% \begin{center}
%  \textbf{\textsc{
% Supplemental note
%  }}
%  \end{center}

\end{document}